\documentclass{aa}
\usepackage{graphicx}
\usepackage{natbib}

\newcommand{\mdot}{\dot{m}}

\begin{document}

\title{A unified accretion-ejection paradigm for Black Hole X-ray Binaries}
\subtitle{I- The dynamical constituents}

\author{Jonathan Ferreira\inst{1}, Pierre-Olivier Petrucci\inst{1}, Gilles Henri\inst{1}, Ludovic
  Saug\'e\inst{1,2} \and Guy Pelletier\inst{1}}
\institute{Laboratoire d'Astrophysique, Observatoire de Grenoble BP53,
  F-38041 Grenoble cedex 9, France
  \and
Institut de Physique Nucleaire de Lyon, 43 bd 11 novembre 1918, F-69622 Villeurbanne cedex, 
France} 

\offprints{J. Ferreira\\ \email{Jonathan.Ferreira@obs.ujf-grenoble.fr}}

%\date{Received .../Accepted ...}
\titlerunning{An accretion-ejection paradigm for BH XrBs}
\authorrunning{Ferreira et al.}

\abstract{We present a new picture for the central regions
of Black Hole X-ray Binaries.  In our view, these central regions have a
multi-flow configuration which consists in (1) an outer standard
accretion disc down to a transition radius $r_J$, (2) an inner magnetized
accretion disc below $r_J$ driving (3) a non relativistic self-collimated
electron-proton jet surrounding, when adequate conditions for pair
creation are met, (4) a ultra relativistic electron-positron beam.

This accretion-ejection paradigm provides a simple explanation to the
canonical spectral states, from radio to X/$\gamma$-rays, by varying the
transition radius $r_J$ and disc accretion rate $\dot m$
independently. Large values of $r_J$ correspond to the
Quiescent state for low $\dot m$ and the Hard state for larger $\dot m$. These states are characterized by the presence of a steady electron-proton MHD jet emitted by the disc below $r_J$. The
hard X-ray component is expected to form at the jet basis. When $r_j$
becomes smaller than the marginally stable orbit $r_i$, the whole disc
resembles a standard accretion disc with no jet, characteristic of the
Soft state. Intermediate states correspond to situations where $r_J \ga
r_i$. At large $\dot m$, an unsteady pair cascade process is triggered
within the jet axis, giving birth to flares and ejection of relativistic
pair blobs. This would correspond to the luminous intermediate state,
sometimes referred to as the Very High state, with its associated
superluminal motions.

The variation of $r_J$ independently of $\dot m$ is a necessary ingredient in this picture. It arises from the presence of a large scale vertical magnetic field threading the disc. Features such as possible hysteresis and the presence of quasi-periodic oscillations would naturally fit within this new framework.

}

\maketitle

\section{Introduction}

Galactic Black Hole X-ray Binaries (hereafter BH XrBs) are binary systems that were first detected as X-ray sources. They harbor a massive compact object as a primary star, being therefore a black hole candidate, accreting matter from the companion.  The X-ray emission probes the inner regions around the compact
object and is interpreted as a signature of accretion (\citealt{tan95c};
\citealt[hereafter McCR03,]{mcc03} and references therein). There is now
growing evidence that these objects also display ejection signatures:
radio emission is commonly interpreted as the presence of compact steady
jets or the sporadic ejection events also seen in infrared and X-ray
bands \citep{fen04b,bux04}. In fact, the correlation between radio
luminosity (ejection) and X-ray luminosity (accretion) found in Active
Galactic Nuclei (AGN) seems to be also consistent with XrBs
\citep{rob04,fal04,gal03,cho03,cor03}. The similarity of the
first detected galactic jet in 1E1740 with extragalactic jets gave birth
to the name "microquasar" \citep{mir98}.\\

What is so dramatic about microquasars is their multiple manifestations
through very different spectral states. They spend most of their
time in the Quiescent state which is characterized by a very low
accretion rate ($\mdot=\dot M_a c^2/L_{Edd}$ as low as $\sim$ 10$^{-9}$). The multiwavelength
spectral energy distributions are thus very scarse but generally show a hard X-ray (2-10 keV)
spectrum with a power law photon index $\Gamma = 1.5 -2.1$ and an optical/UV continuum reminiscent of a $\sim 10^4$ K disc blackbody with strong emission lines (McCR03).\\

Occasionnaly, microquasars enter in outburst, resulting from a dramatic
increase of their accretion rate. During these outbursts, microquasars
show different canonical states, like the well known Hard and Soft 
states. The hard state is characterized by a spectrum dominated above 2
keV by a hard power-law component ($\Gamma \sim 1.5$) with a cut-off around
100 keV (e.g. \citealt{gro98}; \citealt{zdz04} and
references therein), and with a soft X-ray excess below 2 keV interpreted
as the presence of a cool ($\sim$ 0.01-0.2 keV) accretion disc. Strong
radio emission is observed during this state and some VLBI images
directly revealed spatially resolved structures. These are interpreted as non or only mildly relativistic (bulk Lorentz factor $\Gamma_b < 2$) steady jets (e.g. \citealt{sti01,dha00}). On the contrary, the soft
state is dominated by a thermal blackbody-like component, typical of a
standard accretion disc of temperature ranging from 0.7 to 1.5 keV
(consistent with an inner disc radius $r_{in} \sim 10\ r_g= GM/c^2$). A faint power-law component may still be present but with a steep photon index $\Gamma= 2.1-4.8$ (McCR03). This state is
also devoid of any radio emission which is interpreted as the absence of
jet (e.g. \citealt{fen99,cor00}). Hard and soft state span a relatively large range in luminosity, i.e. from 10$^{-2}$ to 1 $L_{Edd}$.\\ 

Microquasars can also be observed in Intermediate states, generally during transitions between the hard and soft states. Intermediate states present relatively
complex spectral and timing behaviors. Detailed studies have been done in
the recent literature (see e.g. \citealt{fen04c,bel05}) and reveal a clear
evolution with time between hard, variable (in X-ray) and radio-loud
systems to softer, less variable and more radio-quiet ones.  The
transitions between the hard and soft "flavors", which seems to
correspond also to a transition between jet-producing and jet-free
states, can be relatively abrupt especially at high luminosity level
where they are apparently coincident with strong radio outbursts
\citep{fen04b}. These non steady ejection events display apparent
superluminal velocities, indicative of a highly relativistic plasma
\citep{mir99,dha00}. These different observational characteristics
lead to the definition of the so-called "jet line" by \citet{fen04b},
that separates hard/jet dominated states to soft/jet-free ones in
hardness-intensity diagrams.

These various spectral states obviously carry a huge amount of
information about the physical processes behind accretion and
ejection. Moreover, since it is believed that microquasars are a scaled
down version of AGN, understanding the various states (along with their
transitions) in XrBs will certainly provide insights on the observed
differences between AGN (radio loud/radio quiet, FRI/FRII, blazars...).\\

The puzzle of the existence of these different spectral states is
enhanced by the fact that each state must correspond to a dynamically
steady state. Indeed, each state lasts a much longer time than the
inferred dynamical time scale. Let us consider the case of GRS1915+105,
which is the BH XrB with the most rapid time scales. It shows states
lasting $\sim 10^3$ sec with rising and decay times of the order of one
second, whereas the keplerian orbit time scale is several milliseconds at
the inner radius \citep{bel97}. One must therefore explain why a system
switches from one stationary state to another one. But before that, one
must first identify the relevant underlying dynamics describing each
state. Up to now, these canonical spectral states are still not fully
understood, let alone the transitions between them.\\ 

The most common paradigm used to interpret observations is based on the low radiative
efficiency of the ADAF model (\citealt{esi97} and references
therein). Within this framework, the highest energy component is due to
the inner thick, low radiative disc whereas an outer standard disc is
responsible for the UV-soft X ray emission. By varying the transition
radius $r_{tr}$ between these two discs (as a function of $\dot m$) one
gets reasonable fits to the spectral energy distributions (hereafter
SEDs, e.g. \citealt{nar96,ham97,esi98,esi01}). In the ADAF paradigm, the accretion power below the transition radius is
essentially stored as thermal energy of protons and eventually advected
below the black hole event horizon. Remarkably, energetic ejection events
appear to be always associated with hard and intermediate states and are
quenched during the soft phase (McCR03 and references therein). This has
led several authors to propose that ADAFs could indeed drive outflows or
its ADIOS extension \citep{bla99,bla04}. However, it must be realized
that there is only one source of energy, namely the release of
gravitational energy through accretion. Hence, whenever a disc is capable
of driving jets, these will carry away a fraction of the released
accretion energy. As a consequence, the disc luminosity will be
quenched. Observations tell us something consistent with this very simple
and unavoidable argument: whenever a steady jet is formed the disc may
{\it observationally} disappear.  This does not mean that the disc is
really disappearing (i.e., inner region being depleted of its mass), only
that we do not see it anymore. Discs that drive jets are indeed radiating only a small fraction of the 
accretion power released as shown in another class of accretion solution, namely Magnetised Accretion-Ejection Structures (hereafter MAES, see e.g. \citealt{fer93a,fer95} and \citealt{fer02} for a review). In this flow, a large scale magnetic field is threading the disc, exerts a torque leading to accretion and allows the production of self-confined jets. A logical consequence is that whenever a jet is
formed, the ADAF hypothesis is no more useful to explain the low
radiative efficiency of the accretion flow.\\

The goal of this paper is to provide an alternative view explaining the
canonical spectral states observed in BH XrBs based on MAES. We
do not intend to address the crucial issue of the transitions between
these states, neither time scales involved nor possible hysteresis
effects. This is delayed for future work. In this paper, we only expose
the global physical picture and show that this framework is rich enough
to explain all known spectral components. In a forthcoming paper, we will
provide calculations of SEDs and compare them to observations.

The paper is then organized as follows. Section~2 describes in some detail the four physical components present within our paradigm, their
dynamical properties as well as radiative processes governing their
emission. The canonical spectral states of BH XrBs are then interpreted within this framework in Section~3. Section~4 is devoted to a discussion of some time scales and timing properties of our model that could be related to some observed features. Section~5 highlights questions opened by our framework as well as future developments.

\section{A novel framework for BH XrBs}

\subsection{General picture}

\begin{figure*}
\includegraphics[width=\textwidth]{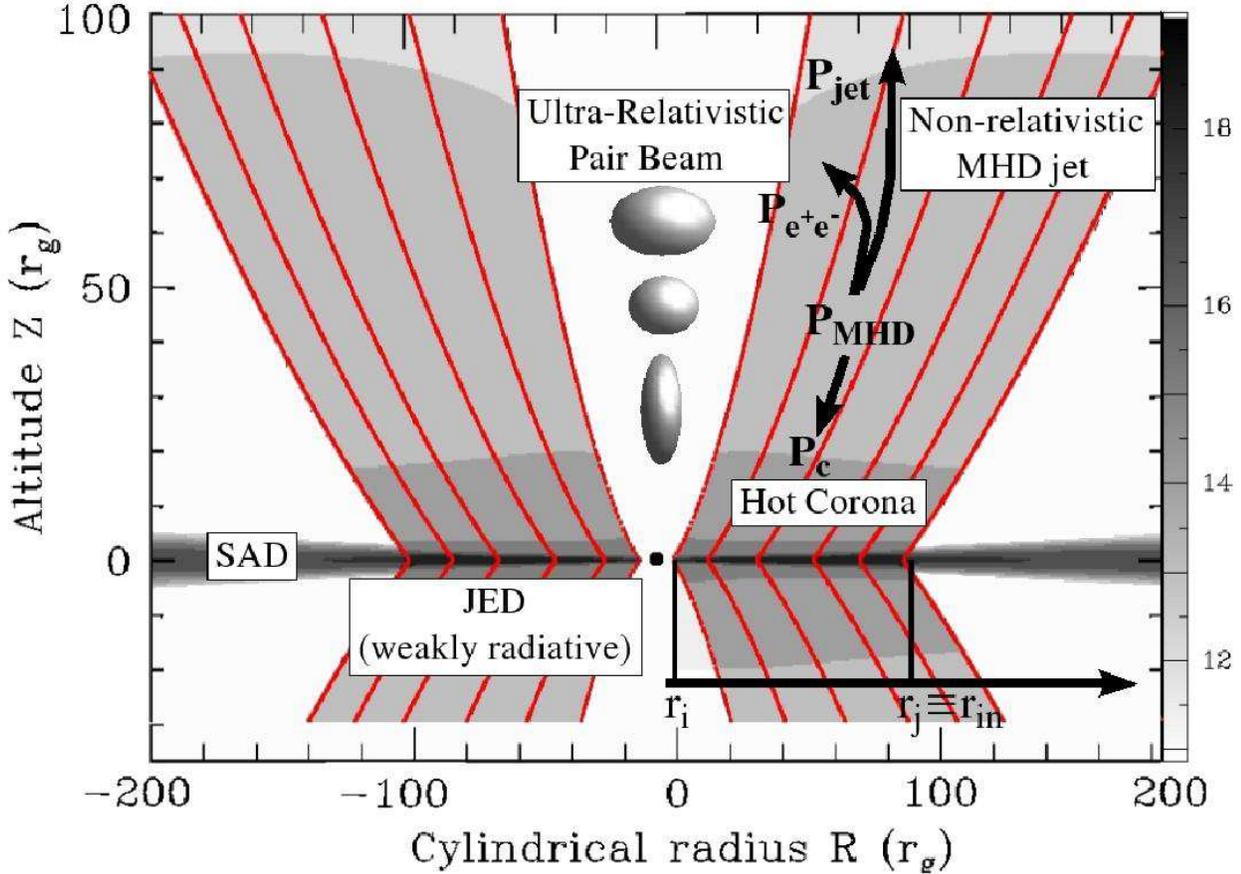}
\caption{A Standard Accretion Disc (SAD) is established down to a radius
$r_J$ which marks the transition towards a low radiative Jet Emitting
Disc (JED), settled down to the last stable orbit. The JED is driving a
mildly relativistic, self-collimated electron-proton jet which, when
suitable conditions are met, is confining an inner ultra-relativistic
electron-positron beam. The MHD power $P_{MHD}$ flowing from the JED acts
as a reservoir for (1) heating the jet basis (radiating as a moving
thermal corona with power $P_c$), (2) heating the inner pair beam
($P_{e^+e^-}$) and (3) driving the compact jet ($P_{jet}$). Field lines
are drawn in solid lines and the number density is shown in
greyscale ($\log_{10} n/\mbox{m}^{-3}$). The MAES solution (JED and MHD jet) was computed
with $\xi=0.01$, $\varepsilon=0.01$ and with $m=10$ and $\dot
m(r_J)=0.01$ (see text).}
\label{fig:smae2flow}
\end{figure*}

We assume that the central regions of BH XrB are composed of four
distinct flows: two discs, one outer "standard" accretion disc (hereafter
SAD) and one inner jet emitting disc (hereafter JED), and two jets, a
non-relativistic, self-confined electron-proton MHD jet and,
when adequate conditions for pair creation are fulfilled, a ultra-relativistic
electron-positron beam. A sketch of our model is shown in
Fig.~\ref{fig:smae2flow} while the four dynamical components are
discussed separately below. This is an extended version of the "two-flow"
model early proposed for AGN and quasars
\citep{pel88b,sol89,pel89,hen91,pel92} to explain the highly relativistic
phenomena such as superluminal motions observed in these sources. This
model provides a promising framework to explain the canonical spectral
states of BH XrBs mainly by varying the transition radius $r_J$ between
the SAD and the JED. This statement is not new and has already been
proposed in the past by different authors
(e.g. \citealt{esi97,bel97,liv03,kin04}) but our model distinguish itself
from the others by the consistency of its disc--jet structure and by the
introduction of a new physical component, the ultra-relativistic
electron-positron beam, that appears during strong outbursts.

%%%%%%%%%%%%%%%
%%%%%%%%%%%%%%%%

We believe that jets from BH XrBs are self-collimated because they follow the same accretion-ejection correlation as in AGN \citep{cor03,fen03,mer03}. This therefore implies the presence of a large scale vertical field anchored somewhere in the accretion disc. We think it is unlikely that such a field has a patchy distribution on the disc. Indeed, once a jet is launched, it exerts a torque on the underlying disc with a corresponding vertical flux of angular momentum and energy. This torque writes $-J_r B_z\sim B_\phi^+ B_z/\mu_o h <0$ where $B_\phi^+$ is the toroidal field at the disc surface. This shows that the product $J_r B_z$ must remain positive over the whole region driving jets. Such a condition is unlikely to be met with a vertical field changing its polarity from one zone to another because the currents (both radial and azimuthal) induced inside the disc will tend to cancel each other. The most efficient way to launch jets from an accretion disc is probably from a radial extension (the JED) with a large scale $B_z$ of the same polarity. 

This vertical field is therefore an unavoidable ingredient to produce self-collimated jets. So, where does this field come from? A first possible origin is in-situ generation of magnetic fields by dynamo. The huge difficulty is to provide an ordered large scale vertical field out of turbulent small scale seed fields. Opening of magnetic loops by the disc differential rotation might provide a mechanism
\citep{rom98}. The second origin is advection of magnetic field by the
accreting material\footnote{Note also that a fraction of the open
poloidal flux initially tied to the primary's progenitor could remain
trapped by the accretion flow.}. We therefore assume that the whole accretion disc is pervaded by a large scale magnetic field $B_z$.

The presence of a large scale vertical field threading the disc is
however not sufficient to drive super-Alfv\'enic jets. This field must be
close to equipartition as shown by \cite{fer95} and \cite{fer97}. The reason is twofold. On one hand, the
magnetic field is vertically pinching the accretion disc so that a
(quasi) vertical equilibrium is obtained only thanks to the gas and
radiation pressure support. As a consequence, the field cannot be too
strong. But on the other hand, the field must be strong enough to
accelerate efficiently the plasma right at the disc surface (so that the
slow-magnetosonic point is crossed smoothly). These two constraints can
only be met with fields close to equipartition.\\

An important local parameter is therefore the disc magnetization $\mu = B_z^2/(\mu_o P_{tot})$ where $P_{tot}$ includes the plasma and radiation pressures. In our picture, a SAD is established down to a radius $r_J$ where $\mu$ becomes of order unity. Inside this radius, a JED with $\mu \sim 1$ is settled. 
At any given time, the exact value of $r_J$ depends on highly non-linear processes such as the interplay between the amount of new large scale magnetic field carried in by the accreting plasma (eg. coming from the secondary) and turbulent magnetic diffusivity redistributing the magnetic flux already present. These processes are far to be understood. For the sake of simplicity, we will treat in the following $r_J$ as a free parameter that may vary with time (see Section~3). 

\subsection{The outer SAD}
\label{SAD}

Accretion requires the presence of a negative torque extracting angular
momentum. In a SAD this torque is assumed to be of turbulent origin and
provides an outward transport of angular momentum in the radial
direction. It has been modeled as an "anomalous" viscous torque of
amplitude $\sim - \alpha_v P_{tot}/r$, where $\alpha_v$ is a small
parameter \citep{sha73}. The origin of this turbulence is now commonly
believed to arise from the magneto-rotational instability or MRI
\citep{bal91}. The MRI requires the presence of a weak magnetic field
($\mu < 1$) and is quenched when the field is close to equipartition. We
make the conjecture that a SAD no longer exists once $\mu$ reaches
unity. We show below that this is very likely to occur in the innermost
regions.

The radial distribution $B_z(r)$ is provided by the induction equation which describes the interplay between advection and diffusion. If we assume that, in steady state, the poloidal field is mostly vertical (no significant bending within the SAD) then this equation writes
\begin{equation}
\nu_m \frac{\partial B_z}{\partial r} \simeq u_r B_z
\end{equation}
where $\nu_m$ is the turbulent magnetic diffusivity. This equation has the obvious  {\it exact} solution 
\begin{equation}
B_z \propto  r^{- {\cal R}_m}
\label{eq:Bz}
\end{equation}
where ${\cal R}_m = -r u_r/\nu_m$ is the (effective) magnetic
Reynolds number. In a turbulent disc one usually assumes that all anomalous transport
coefficients are of the same order so that $\nu_m \simeq \nu_v$, $\nu_v$
being the turbulent viscosity. Since the (effective) Reynolds number
${\cal R}_e = -r u_r/\nu_v = 3/2$ in a SAD, one gets that any vertical
magnetic field is naturally {\it increasing} towards the center. Now in a
SAD of vertical scale height $h(r) \propto r^\delta$, the total pressure
$P_{tot} = \rho \Omega_k^2 h^2 $ ($\Omega_k$ is the Keplerian rotation
rate) scales as
\begin{equation}
P_{tot} = \frac{\dot M_a \Omega_k^2 h}{6\pi \nu_v} \propto r^{-3/2 - \delta} 
\end{equation}
where $\dot M_a$ is the (constant) disc accretion rate. Using Eq.~(\ref{eq:Bz}) we get 
\begin{equation}
\mu \propto r^{-\epsilon} \, \, \mbox{      with      }\, \,  \epsilon = 2 {\cal R}_m - \delta - 3/2 
\end{equation}
In a SAD ${\cal R}_m \simeq 3/2$ and $\delta$ is always close to unity (apart from the unstable radiation pressure dominated zone where $\delta=0$). Of course, the real value of $\epsilon$ critically depends on the magnetic Prandtl number (${\cal P}_m = \nu_v/\nu_m$) but this result suggests that one may reasonably expect $\mu$ to increase towards the center. Whenever a BH XrB reaches $\mu \simeq 1$ at a radius $r_J > r_i$, $r_i$ being the last marginally stable orbit, the accretion flow changes its nature to a JED.
 
To summarize, the accretion flow at $r> r_J$ is a SAD with  $\mu \ll 1$ fueled by the companion's mass flux and driving no outflow (constant accretion rate $\dot M_a$). The global energy budget is $P_{acc,SAD} = 2 P_{rad,SAD}$ where
\begin{equation}
\label{paccsad}
P_{acc,SAD}  \simeq \frac{GM \dot M_a}{2r_J}
\end{equation}
and $P_{rad,SAD}$ is the disc luminosity (from one surface only). Its emission has the characteristic multi-blackbody shape produced by a radial temperature distribution $T_{eff}(r) \propto r^{-3/4}$. The spectrum is therefore dominated by the hottest inner parts at $r_J$.

\subsection{The inner JED}
\label{JED}

This inner region with $\mu \sim 1$ is fueled by the SAD at a rate $\dot
M_{a,J} = \dot M_a (r_J)$. Since it undergoes mass loss, 
the JED accretion rate is written as
\begin{equation}
\dot M_a(r) = \dot M_{a,J} \left (\frac{r}{r_J} \right )^\xi
\end{equation}
where $\xi$ measures the local ejection efficiency \citep{fer93a}. The global energy
budget in the JED is $P_{acc,JED} = 2 P_{rad,JED} + 2 P_{MHD}$ where
$P_{MHD}$ is the MHD Poynting flux feeding a jet, whereas the
liberated accretion power writes
\begin{equation}
\label{paccjed}
P_{acc,JED} \simeq \frac{GM \dot M_{a,J}}{2r_i} \left [ \left (
\frac{r_i}{r_J} \right )^\xi - \frac{r_i}{r_J} \right ]
\label{eq:bilan}
\end{equation}
The dynamical properties of a JED have been extensively studied in a
series of papers (see \citealt{fer02} and references therein). We will
here only briefly recall the main properties and refer the interested
reader to specific papers. In this dynamical structure accretion and
ejection are interdependent: jets carry away the exact amount of angular
momentum allowing the disc material to accrete. The ratio at the disc
midplane of the jet torque to the turbulent "viscous" torque is
\begin{equation}
\Lambda \sim \frac{ B_\phi^+ B_z/\mu_o h}{\alpha_v P_{tot}/r} \sim \frac{B_\phi^+ B_z}{\mu_o P_{tot}} \frac{r}{\alpha_v h}
\end{equation}
It is straightforward to see that the necessary condition to drive jets
(fields close to equipartition) from Keplerian discs leads to a dominant
jet torque. In fact, it has been shown that steady-state ejection
requires $\Lambda \sim r/h \gg 1$ \citep{fer97,cas00a}.

This dynamical property has a tremendous implication on the JED
emissivity. The JED luminosity comes from the accretion power dissipated
within the disc by turbulence and transported away by photons, so
$2P_{rad,JED}= P_{diss}$. This dissipated power is very difficult to
estimate with precision because it requires a thorough description of the
turbulence itself. Thus, one usually uses crude estimates based on
"anomalous" turbulent magnetic resistivity $\eta_m$ (Joule heating) and
viscosity $\eta_v$ (viscous heating). This translates into $P_{diss}=
P_{Joule} + P_{visc} = \int \eta_m J^2 dV + \int \eta_v (r\partial \Omega
/\partial r)^2 dV$ where integration is made over the whole volume
occupied by the JED.  The importance of local "viscous" dissipation with
respect to the MHD Poynting flux leaving the disc is approximately given
by
\begin{equation}
\frac{P_{visc}}{2P_{MHD}}\sim \frac{1}{\Lambda} 
\end{equation}
which is much larger than unity: turbulent "viscosity" provides negligible dissipation in a JED. Joule heating arises from the dissipation of toroidal and radial currents which are comparable\footnote{Full computations of MAES show that the three magnetic field components are
comparable at the disc surface, namely $B_\phi^+ \sim B_r^+ \sim B_z$
\citep{fer95,fer97}.}. One therefore gets $\eta_m J^2 \sim \nu_m B_z^2/\mu_o h^2 \sim \nu_v \rho \Omega^2 \sim \eta_v (r \partial \Omega/\partial r)^2$, for equipartition fields, isotropic magnetic resistivity $\eta_m = \mu_o \nu_m$ and a turbulent magnetic Prandtl number of order unity. This leads to
\begin{equation}
\frac{P_{Joule}}{2 P_{MHD}} \sim \frac{1}{\Lambda}
\end{equation}
namely a negligible effective Joule heating. Thus, the total luminosity $2P_{rad,JED}$ of the JED is only a fraction $1/(1+\Lambda)$ of the accretion disc liberated power $P_{acc,JED}$.\\ 

To summarize, under quite general conditions on the turbulence within
magnetized discs, most of the available accretion energy is powering the
outflowing plasma \citep{fer93a,fer95}. This is in strong contrast with
ADAFs where the accretion power is stored as heat advected by the
accreting plasma. In this case, low luminosity discs can be obtained as
long as the central object possesses an event horizon. However, the power
to magnetically drive jets is also missing. In the case of MAES, the JED
is weakly dissipative while powerful jets are being produced regardless
of the nature of the central object.
 
Complete calculations of MAES showed that isothermal or adiabatic
super-Alfv\'enic jets from Keplerian accretion discs were possible only
when a tiny fraction of the accreted mass is locally ejected. This
translates into a small ejection efficiency, typically $\xi \sim 0.01$
\citep{fer97,cas00a}.  When some heat deposition occurs at the JED upper
layers a typical value of $\xi \sim 0.1$ becomes possible, even up to 0.5
but never reaching unity, in agreement with Eq.~(\ref{eq:bilan})
\citep{cas00b}. This is a much lower mass loss than that assumed in ADIOS
models \citep{bla99}.\\

The fact that the jet torque largely dominates the turbulent torque provides another striking difference between the internal structures of SADs and JEDs. Indeed, the angular momentum conservation provides a sonic Mach number measured at the disc midplane
\begin{equation}
m_s = - \frac{u_r}{C_s} = \alpha_v \varepsilon (1 + \Lambda)
\label{eq:ms}
\end{equation}  
\noindent where $C_s= \Omega_k h$ is the sound speed. Thus, a SAD displays $m_s = \alpha_v \varepsilon \ll 1$ whereas a JED has a much higher accretion velocity, namely $m_s \simeq 1$ \citep{fer95,fer97}. This has two major consequences. First, a JED is much less dense than a SAD\footnote{It has been recently showed that a previously claimed instability of accretion-ejection structures does not apply to this type of solution (see \citet{kon04} and references therein).}. Second, there is a stronger bending of the poloidal field lines. Indeed, in spite of the same turbulent magnetic diffusivity ($\nu_m \sim \nu_v$), the larger accretion velocity $u_r$ leads to an effective magnetic Reynolds number ${\cal R}_m \sim \varepsilon^{-1}$ where $\varepsilon= h/r$ is the disc aspect ratio \citep{fer95}. This translates into a field at the disc surface verifying $B_r^+/B_z \sim {\cal R}_m \varepsilon \ga 1$, as required to magnetically launch cold jets.

Mass conservation in the JED writes
\begin{eqnarray}
n &=& \frac{\dot M_a(r)}{4 \pi m_p \Omega_k r^3} m_s^{-1} \varepsilon^{-2} \nonumber \\
  &\simeq & 10^{25}\ \varepsilon^{-2} \dot m m^{-1} R^{\xi -\frac{3}{2}} \ \mbox{m$^{-3}$} 
\end{eqnarray}
where $m= M/M_{\sun}$, $R= r/r_g$ ($r_g=GM/c^2$) and $\dot m = \dot M_{a,J} c^2/L_{Edd}$. This density requires a magnetic field
\begin{eqnarray}
B_z &= & \left ( \frac{\mu}{m_s}\right )^{1/2}  \left ( \frac{\mu_o \dot M_a(r) \Omega_k}
{4 \pi r}\right )^{1/2}  \nonumber \\
& \simeq& 4.4\ 10^{8}\ \dot m^{1/2} m^{-1/2} R^{\frac{\xi}{2}-\frac{5}{4}}\ \mbox{G} 
\label{Bz}
\end{eqnarray}
For illustration, we provide below the case of an optically thick, Thomson dominated and gas pressure supported JED. In this region, the disc aspect ratio is
\begin{equation}
\varepsilon = h/r = 2.6\ 10^{-3} \ \dot m^{1/4} m^{-1/8} R^{\frac{\xi}{4}+\frac{1}{16}}   
\end{equation}
which allows to precisely specify the above quantities. Moreover, the ratio of radiation to gas pressure, Thomson opacity, effective and central temperatures are
\begin{eqnarray}
\frac{P_{rad}}{P_{gas}} & =& 0.3\ \dot m R^{\xi-1} \\
\tau_T & \simeq & n \sigma_T h = 3.8\ 10^2\ \dot m^{3/4} m^{1/8} R^{\frac{3\xi}{4}-\frac{9}{16}} \\
T_{eff} & \simeq & 876 \ \dot m^{5/16} m^{-9/32} R^{\frac{5\xi}{16}-\frac{47}{64}} \ \mbox{eV} \\ 
T_o &=& 3.2\ \dot m^{1/2} m^{-1/4} R^{\frac{\xi}{2}-\frac{7}{8}}\ \ \ \mbox{ keV}
\end{eqnarray}
The spectrum emitted by an optically thick JED is a multi-blackbody but
with a temperature drastically reduced from that of a SAD. As a
consequence, the flux emitted by the JED is expected to be unobservable
with respect to that of the outer SAD. In practice, this mostly depends
on the radial extension of the JED. Indeed, the flux emitted by the
surrounding SAD scales as $T_{eff,SAD}^4 r^2$ measured at $r_J$ whereas
the flux emitted by the JED is dominated by $T_{eff,JED}^4 r^2$ measured
at $r_i$. One therefore gets that the ratio of the JED to the SAD flux
scales as $(r_J/r_i)/(1+\Lambda) \sim \varepsilon r_J/r_i$ which is much
smaller than unity for reasonable values of $r_J$.  Thus, the values of
the "disc inner radius" ($r_{in}$) and "disc accretion rate"
observationally determined from spectral fits must be understood here as
values at the transition radius, namely $r_{in} \equiv r_J$ and $\dot m
\equiv \dot m(r_J)$: the optically thick JED is spectrally hardly
visible.

%%%%%%%%%%%%%%%%%%%%%%%%%%%%%%%%
\subsection{Non-relativistic electron-proton jets from JEDs}
\label{MAES}

The ejection to accretion rate ratio in a JED writes $2 \dot M_{jet}/\dot
M_{a,J} \simeq \xi \ln (r_J/r_i)$. In principle, the ejection efficiency
$\xi$ can be observationally deduced from the terminal jet speed. Indeed,
the maximum velocity reachable along a magnetic surface anchored on a
radius $r_o$ (between $r_i$ and $r_J$) is $u_\infty \simeq \xi^{-1/2}
\sqrt{GM/r_o}$ in the non-relativistic limit (see \citealt{fer97} for
relativistic estimates). Although a large power is provided to the
ejected mass (mainly electrons and protons), the mass loss ($\xi$) is
never low enough to allow for speeds significantly relativistic required
by superluminal motions: MHD jets from accretion discs are basically non
or only mildly relativistic with $u_\infty \sim 0.1 - 0.8\ c$
\citep{fer97}. This is basically the reason why they can be efficiently
self-confined by the magnetic hoop stress. Indeed, in relativistic flows
the electric field grows so much that it counteracts the confining effect
due to the toroidal field. This dramatically reduces the self-collimation
property of jets \citep{bog01,bog01b,pel04}.\\

Calculations of jets crossing the MHD critical points have been undergone
under the self-similar ansazt \citep{cas00b,fer04}. In these
calculations, the emission of the MHD jet has been neglected and all the
available power is converted into ordered jet kinetic energy. However, a
fraction of this power is always converted into heat and particle
acceleration, leading to emission. In our case, jets from MAES have two
distinct spectral components and the resultant SED may therefore be quite
intricate. Producing a global SED is out of the scope of the present paper. It requires to fix several parameters, which is legitimate only by object fitting. This is postponed  for future work.

\subsubsection{A non-thermal extended jet emission}

We expect a small fraction of the jet power $P_{jet}$ to be converted
into particles, through first and/or second order Fermi acceleration,
populating the MHD jet with supra-thermal particles. These particles are
responsible for the bulk emission of the MHD jet. This is similar to
models of jet emission already proposed in the literature
\citep{fal95a,vad01,mark01,mark03,mark04b,fal04}. 
In these models, the jet is assumed to be radiating self-absorbed
synchrotron emission in the radio band becoming then optically thin in
the IR-Optical bands and providing a contribution up to the X/$\gamma$-rays.
A flat or even inverted spectrum index in the radio band is quite easily achieved by self-collimated jets for reasonable values of the parameters (e.g. the exponent $p$ of the power law particle distribution).

Note that the MHD jet due to the MAES yields $B \propto m^{-1/2} \dot m^{1/2}$. In the framework of \citet{hei03}, this implies that such a jet would provide correlated radio and X-ray emissions close to the observed law, namely $F_R \propto F_X^{0.7}$ \citet{gal03,cor03}. However, the spectrum index in the X-ray band would not be steep enough, even taking into account the cooling of the particles. Moreover, the fundamental plane of BH activity of \citet{mer03}, namely the correlation between mass, radio and X-ray fluxes, cannot be explained by such synchrotron jets \citep{hei04}. Following these authors, we conclude that there must be another significant contribution to the X-ray emission in the Low/hard state. The same conclusion was independently reached by
\cite{rod04} who suggested that, to explain the energy dependence of the
quasi-periodic oscillation (QPO) amplitude in GRS 1915+105, the high
energy spectrum of the source must be the sum of different emission
processes. Another argument comes from the study of the overall spectral energy distributions. 
At least in some objects, the extrapolation of the X-ray power-law spectrum towards the optical and infrared bands is above the observed fluxes, which shows that hard X-rays cannot be direct synchrotron radiation from  the jet \citep{kale05}.  

Moreover, BH XrBs in the hard state generally exhibit a spectrum with a
high energy cut-off around 100 keV (e.g. \citealt{gro98}; \citealt{zdz04}
and references therein). While naturally obtained if the emission is
thermal (i.e. a comptonized corona), non-thermal emission requires a fine
tuning of the parameters that we found questionable. However, our
framework naturally provides another contribution to the high energy
emission as long as a JED is present.

\subsubsection{A thermal jet basis}

Jet production relies on a large scale magnetic field anchored on the
disc as much as on MHD turbulence triggered (and sustained) within
it. This implies that small scale magnetic fields are sheared by the disc
differential rotation, leading to violent release of magnetic energy at
the disc surface and related turbulent heat fluxes (e.g. \citealt{gal79,
hey89a,sto96, mer02}). The energy released is actually tapping the MHD
Poynting flux flowing from the disc surface. We can safely assume that a
fraction $f$ of it would be deposited at the jet basis, 
with a total power $P_c = f P_{MHD}$. The dominant cooling term in this
optically thin medium is probably comptonization of soft photons emitted
by the outer SAD (with a small contribution from the underlying JED). These are
circumstances allowing a thermal plasma to reach a temperature as high as
$\sim 100$ keV, \citep{kro95,mah97,esi97}.\\

The computation of the exact spectral shape produced by this "corona"
through thermal comptonization requires sophisticated computations
(e.g. \citealt{haa93,pou96}) which are out of the scope of this
paper. Instead, a cut-off power law shape is generally used as zero-order
approximation. In this case, the high energy cut-off is rougly equal to
twice the plasma temperature (see e.g. \citealt{pet00} for more
discussion). The power law photon index can also be approximated by a
simple fonction of the Compton amplification factor $A$, which is equal
to the ratio of the total luminosity outgoing from the jet basis to the soft
luminosity $P_{soft}$ entering in it (see e.g. \citealt{bel99,mal01}): 
\begin{equation}
\label{eq:gamma}
\Gamma \simeq C(A-1)^{-\eta} \mbox{    with    } A=1+ P_c/P_{soft}
\end{equation}
where $C$ and $\eta$ depend on the geometry of the disc-corona
configuration.\\

In the most general case, $P_{soft}$ should include the SAD and JED
emissions but also the reprocessed radiation from the discs (both JED and SAD)
that are partly intercepted by the corona. In consequence, the photon
index depends implicitly on parameters like $f$, $\Lambda$, $r_J$ but
also on the bulk motion of the corona in a complex manner. This will be
precisely discussed in a forthcoming paper where calculations of SEDs will
be provided. Simple estimates can be given, however, in the case of large
$r_J$ and $\Lambda$, since in these conditions the SAD and JED emissions
become negligible compared to the reprocessed one. This situation has
been precisely studied by \cite{mal01}. The parameters $C$ and $\eta$ of
Eq.~\ref{eq:gamma} obtained by these authors are equal to 2.19 and 2/15
respectively. Hard X-rays photon indexes in the range 1.4-2 are easily obtained
for different corona velocities and corona aspect ratios
(i.e. height/width). These values are in good agreement with what is
generally observed in the hard states where we expect large $r_J$
(cf. Sect.~\ref{specstate} for a more detailed discussion). We can note
also that a decrease of $r_J$ will result in a larger $P_{soft}$, due to
the increase of the SAD emission, and thus in a softening of the X-ray
spectrum.

%%%%%%%%%%%%%%%%%%%%%%%%%%%%%%%%%%
\subsection{The inner ultra-relativistic pair beam}
%%%%%%%%%%%%%%
\label{pairbeam}

Since the large scale magnetic field driving the self-confined jet is
anchored onto the accretion disc which has a non zero inner radius, there
is a natural hole on the axis above the central object with no baryonic outflow (this also holds for neutron stars). This hole provides a place for pair
production and acceleration with the outer MHD jet acting as a sheath
that confines and heats the pair plasma. This is the microquasar version
of the "two flow" model that has been successfully applied to the high
energy emission of relativistic jets in AGNs
\citep{hen91,mar95,mar98,ren98}. \\

The $e^+-e^-$ plasma is produced by $\gamma-\gamma$ interaction, the
$\gamma$-ray photons being initially produced by a few relativistic
particles by Inverse Compton process, either on synchrotron photons (Synchrotron Self Compton or SSC) or on disc photons (External Inverse Compton or EIC). Detailed models for AGNs have shown that all processes can contribute, depending on the physical parameters of the system (magnetic field, disc luminosity, distance). We do not intend to build an explicit Spectral Energy Distribution of the pair plasma here, but we will just discuss the general mechanism of pair beam formation and the relevance of each process in explaining the various non thermal components.\\

It is well known that above 0.5 MeV photons can annihilate with themselves to produce an electron-positron pair. Usually, pairs are assume to cool once they are formed, producing at turn non thermal radiation. Some of this radiation can be absorbed to produce new pairs, but the overall pair yield never exceeds 10 \%.  A key point of the two-flow model however is that the MHD jet launched from the
disc can carry a fair amount of turbulent energy, most probably through
its MHD turbulent waves spectrum. A fraction of this power can
be transferred to the pairs ($P_{e^+e^-} << P_{MHD}$). Thus the freshly created pairs can be continuously reheated, triggering an efficient pair runaway process leading to a dense pair plasma \citep{hen91}.\\

As we said, reacceleration is balanced by cooling through the combination of synchrotron, SSC and EIC processes. Synchrotron and SSC emission are quasi isotropic in the pair frame, but the external photon field is strongly anisotropic. The pair plasma will then
experience a strong bulk acceleration due to the recoil term of EIC, an effect
also known as the "Compton Rocket" effect \citep{ode81,ren98}. As was shown in previous works, this "rocket" effect is the key process to explain relativistic motions \citep{mar95, ren98}.\\

 At a given distance of the disc, 
the bulk acceleration saturates at a characteristic Lorentz factor, depending only on 
the radiation angular distribution. It is defined by the condition that the net radiation flux in the comoving frame (after Lorentz transform) vanishes. It can be shown easily that this characteristic Lorentz factor is approximately $\Gamma_{b,eq} \simeq (z/r_i) ^{1/4}$ on the axis of a  standard accretion disc \citep{ren98}, as long as the distance $z$ verifies $r_J\ll z $: here the relevant disc inner radius is in fact the transition radius $r_J$. Noticeably this value does not depend on the disc luminosity (or accretion rate): it is only dependent on the angular distribution of the intensity, i.e. the radial dependency of the temperature $T \propto r^{-3/4}$.  The photon field is in fact dominated by photons emitted at $r \sim z$. The modification introduced by the JED in the central region is likely to be immaterial for two reasons. First although the luminosity decreases, the radial dependency remains almost unchanged. Second, as we argue below, the final bulk Lorentz factor depends only on the distance where the plasma decouples from the radiation, and not on the motion close to the core.\\
A pure pair plasma will thus experience  a continuous bulk acceleration, the bulk velocity increasing slowly with the distance. At some point, the radiation field becomes too weak to act efficiently: this happens when the relaxation time towards the equilibrium Lorentz factor becomes larger than the dynamical time $z/c$. At this distance, the acceleration process stops and the plasma decouples from the external radiation field, moving on at a ballistic constant Lorentz factor $\Gamma_{b,\infty}$. The asymptotic Lorentz factor depends essentially on the location of this critical distance.\\
 
In the original O'Dell's version of this process, the pair plasma was not
supposed to be reheated and this effect has been shown to be rather
inefficient, because cooling is always much faster than acceleration
\citep{phin87}. In fact, for a cold plasma, the mechanism reduces to the ordinary radiation pressure. Under these conditions, the critical distance is approximately $\ell_s^{4/7}$,where 
 $\ell_s =  \sigma_T P_{rad,SAD}/ 4 \pi m_e c^3 r_J$ is the soft photon compactness. For a near Eddington accreting disc, $\ell_s \simeq 10^3 $ and  $\Gamma_{b,\infty} \sim \ell_s^{1/7} \sim 2 -3$ \citep{phin87}. Although this is indeed a relativistic motion (an apparent superluminal motion is possible), this may not be high enough to account for high values around 5, as observed in microquasars. \\
 
In the two flow model however, continuous reheating of the pairs makes
the bulk acceleration more efficient, acting thus over a much larger distance : the radiation force is multiplied by $<\bar{\gamma}^2>$, where $\bar{\gamma}$ is the random (or relativistic temperature) of the pair plasma, (not to be confused with the bulk Lorentz factor $\Gamma_b$). Although the equilibrium Lorentz factor at a given distance is unchanged, the critical decoupling distance is much larger. The asymptotic 
bulk Lorentz factor becomes  $(\ell_s <\bar{\gamma}^2>/<\bar{\gamma}>)^{1/7}$ and values of 5 to 10 can be easily reached in near-Eddington accretion regime around stellar black
holes \citep{ren98}.\\

Producing this pair plasma requires thus altogether a strong MHD jet, a
radiative non-thermal component extending above the MeV range and a minimal
$\gamma-\gamma$ optical depth, namely $\tau_{\gamma \gamma} \sim 1$. The non thermal component can indeed be associated with the steep power law observed during the intermediate states, given the fact that it seems to extend to MeV range without any break (McCR03). It is most probably due to Inverse Compton process on the disc photons. Indeed, the optical depth $\tau_{\gamma \gamma}$ for absorbing photons with energy $E_\gamma= \varepsilon m_{\rm e} c^2$ is approximately for a
spherical source of radius $R$ filled by soft photons with density
$n(\varepsilon)$ by unit reduced energy:
 \begin{equation}	
\tau_{\gamma \gamma} = \frac{1}{\varepsilon} n(\frac{1}{\varepsilon}
)\sigma_T R = \frac{\sigma_T \varepsilon} {4 \pi m_{\rm e} c^3 R} (\nu
L_{\nu})_{m_{\rm e} c^2/h\varepsilon }
\end{equation}
We take a typical soft power-law spectrum $\nu L_{\nu} = E L_E = L_0
(E/E_0)^{-\Gamma+2}$ where $\Gamma$ is the soft photon index, typically
around 2.5 for luminous intermediate states.  Adopting this nominal value, the
$\gamma-\gamma$ optical depth becomes a fonction of the energy
$E_\gamma$:
\begin{eqnarray}
\tau_{\gamma \gamma} (E_\gamma) &=& 0.7\times 262^{(2.5-\Gamma)} \left(
\frac{L_0}{0.1 L_{Edd}} \right) \left(\frac{R}{30 r_g}\right)^{-1}\\ &&
\times \left(\frac{E_0}{1 \rm{ keV}}\right)^{\Gamma-2}
\left(\frac{E_\gamma}{1 \rm{ MeV}}\right)^{\Gamma-1} \nonumber
\label{eqtaugg}
\end{eqnarray}
Thus, assumptions that appear quite reasonable for
the luminous intermediate state provide good conditions
for pair creation.\\
 
It is noteworthy that the pair beam is intrinsically highly variable and
subject to an intermittent behavior. Indeed, once the pair creation is
triggered, a regulation mechanism must occur to avoid infinite power of
the pair plasma and limit the pair run-away. This is probably accomplished by
the quenching of the turbulence ($P_{e^+e^-}$ vanishes) when most of its
energy is suddenly tapped by the catastrophic number of newly created
pairs. These pairs will therefore simply expand freely, confined by the
heavier MHD jet. One would then expect a flare in the compact region,
followed by the ejection of a superluminal radio component, analoguous to those observed in AGNs (Saug\'e \& Henri 2005, A\&A submitted). Such a situation can repeat itself as long as the
required physical conditions are met. Alternatively, it may also that the formation of a dense pair beam destroys the surrounding MHD jet, explaining the disappearance of the compact jet after a strong ejection event.

%%%%%%%%%%%%%%%%%%%%%%%%%%%%%
\section{Canonical spectral states of X-ray binaries}
\label{specstate}

\subsection{The crucial roles of $r_J$ and $\mdot$}

From Section~2, it is clear that the spectral appearance of a BH XrB critically depends
on the size of the JED relative to the SAD, namely $r_J$. As stated before, $r_J$ is the 
radius where the disc magnetization $\mu = B_z^2/(\mu_o P_{tot})$ becomes of order unity. 
Thus, $r_J$ depends on two quantities $P_{tot}(r,t)$ and $B_z(r,t)$. The total pressure is directly proportional to $\mdot$ since $P_{tot}= \rho \Omega_k^2 h^2
\propto \mdot m^{-1}r^{-5/2}$. As a consequence, any variation of the
accretion rate in the outer SAD implies also a change in the amplitude of the
total pressure. But we have to assume something about the time evolution of the large scale magnetic field threading the disc. {\em Within our framework, $\dot m$ and $B_z$ are two quantities that may vary independently with time}. \\

Let us assume that $B_z$ does not depend on the disc accretion rate. If, for instance, $\dot m$ undergoes a sudden increase (triggered by, e.g. some disc instability at the outer radii \citealt{las96}),
then there is a corresponding increase of $P_{tot}$ which is propagating
inwards, eventually reaching the inner JED. Since no magnetic flux is being
simultaneously added, the region where the vertical magnetic field
is close to equipartition shrinks, namely $r_J$ decreases. If, on the
contrary, the disc accretion rate decreases (without a corresponding
decrease of the magnetic flux threading the disc), then the decrease of
$P_{tot}$ requires some diffusion of the vertical magnetic field in the
inner regions in order to maintain equipartition, hence the JED expands and $r_J$ increases.
According to this simplistic argument, one would expect an anti-correlation between $r_J$ 
and $\dot m$, namely  a larger $\mdot$ implies a smaller $r_J$ (and vice-versa).  

Alternatively, one could also argue that a larger $\mdot$ implies more plasma within the disc and that a larger $B_z$ would then be locally generated by dynamo. If such a process provides $B_z \propto \dot m^{1/2}$, then $r_J$ would always remain unchanged, whatever $\dot m$. Another alternative could be advection of the companion's magnetic field along with the flow. Now, because of the stellar dynamo, such a field could also change with a time scale very different from that related to changes in $\dot m$. In any case, the amount and polarity advected along strong accretion phases would be an unknown function $B_z(\dot m)$.\\

The processes governing the amplitude and time scales of these
adjustments of $r_J$ to a change in $\mdot$ are far too complex to be
addressed here. They depend on the nature of the magnetic diffusivity
within the disc but also on the radial distribution of the vertical magnetic field. We will
simply assume in the following that $r_J$ and $\dot m$ are two independent parameters. In that respect, our view is very different from that of \cite{esi97,mah97}  who considered only the dependency of $\dot m$ to explain the different spectral states of BH XrBs.

\begin{figure*}
\begin{center}
\includegraphics[height=12cm]{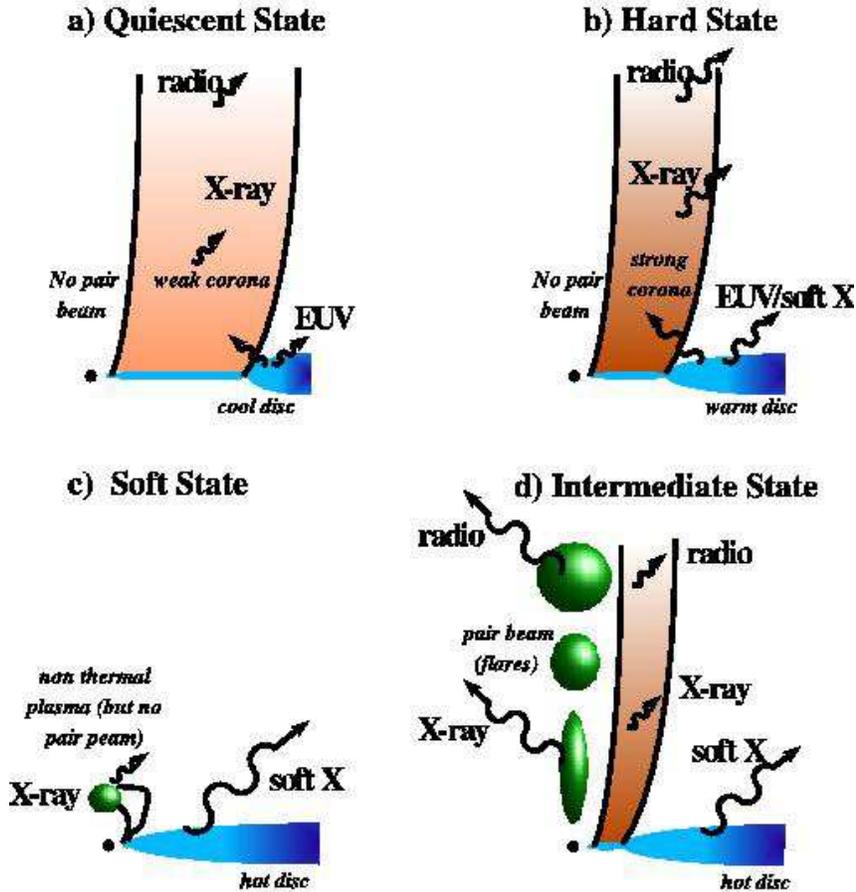}
\end{center}
\caption{The canonical spectral states of BH X-ray binaries. {\bf (a)}
Quiescent state obtained with a low $\mdot$ and a large
$r_J$: the Jet Emitting Disc (JED) occupies a large zone in the accretion
disc. {\bf (b)} Hard state with much larger $\mdot$ and smaller $r_J$:
the pair creation threshold is still not reached. {\bf (c)} Soft state when $\mdot$ is such that there is no zone anymore within the disc where an equipartition field is present: no JED, hence neither
MHD jet nor pair beam. {\bf (d)} Luminous Intermediate state between the Hard and the Soft states: the high disc luminosity (SAD) combined with the presence of a MHD jet allows pair creation and
acceleration along the axis, giving birth to flares and superluminal
ejection events.}
\label{diffstate}
\end{figure*}

\subsection{The Quiescent state} 

This state is characterized by a very low accretion rate ($\mdot$ as low
as $\sim$ 10$^{-9}$) with a hard X-ray component. The ADAF model has been successfully applied to
some systems with a large transition radius between the ADAF and the
outer standard disc, namely $r_{tr} \sim 10^3-10^4\ r_g$
(e.g. \citealt{nar96,ham97}). However, such a model does not account for jets
and their radio emission, even though XrBs in quiescence seem also to
follow the radio/X-ray correlation (e.g. \citealt{fen03,gal04,gal05}).

Within our framework, a BH XrB in quiescence has a large $r_J$, so that a large zone in the whole disc is driving jets (Fig.~\ref{diffstate}a). The low $\dot m$ provides a low synchrotron jet luminosity, while the JED is optically thin, producing a SED probably very similar to that of an ADAF. We thus expect $r_J \sim r_{tr}$. The weak MHD Poynting flux prevents the ignition of the pair cascade process and no pair beam is produced.

It must be noted that slightly more complicated situations can arise depending on the actual value of $\mdot$. For instance, the innermost denser regions of the JED could become optically thick for larger values of $\mdot$, i.e closer to the Hard state level.

\subsection{The Hard state} 

Within our framework, the JED is now more limited radially than in the
Quiescent state, namely $r_J \sim 40-100\ r_g$
(Fig.~\ref{diffstate}b). This transition radius corresponds to the inner
disc radius $r_{in}$ as obtained within the SAD framework
\citep{zdz04b}. Due to the higher $\dot m$, the JED may become optically
thick, which is required to explain the broad iron lines observed in some
binary systems \citep{now02,sid02,fro01}. The low velocity of the plasma
expected at the jet basis is in good agreement with recent studies of
XrBs in Hard state \citep{mac03b,gal03}. It can also explain the apparent
weakness of the Compton reflection \citep{zdz99,gil99} as already
suggested by \citet[see also \citealt{bel99,mal01}]{mark03} and tested by
\cite{mark04}. In any case, the JED intrinsic emission is weak with
respect to that of the outer standard disc: most of the accretion power
flows out of the JED as an MHD Poynting flux. Nevertheless, the threshold
for pair creation is still not reached and there is no pair beam, hence
no superluminal motion. The MHD power is therefore shared between the jet
basis, whose temperature increases (the thermal "corona") producing
X-rays, and the large-scale jet seen as the persistent (synchrotron)
radio emission.

\subsection{The Soft state} 

Our interpretation of the Soft state relies on the disappearance of the
JED, i.e. when $r_J$ becomes smaller or equal to $r_i$ (Fig.~\ref{diffstate}c). Depending on the importance of the magnetic flux in the disc, this may occur at different accretion rates. Thus, the threshold in $\dot m$ where there is no region anymore in the disc with equipartition fields may vary. The
whole disc adopts therefore a radial structure akin to the standard disc
model. Since no MHD jet is produced, all associated spectral signatures
disappear. Even if pair production may take place (when $\dot m$ is large), the absence of the confining MHD jet forbids the pairs to get warm enough and be accelerated: no superluminal motion should be
detected.

Note however that the presence of magnetic fields may be the cause of particle
acceleration responsible for the weak hard-energy tail (McCR03, \citealt{zdz04} and
references therein).

\subsection{Intermediate states}
 
This state has been first identified at large luminosities ($L> 0.2\
L_{Edd}$) and was initially called Very High state. However, high
luminosity appeared to not be a generic feature since it has be observed at
luminosities as low as $0.02\ L_{Edd}$ (McCR03, \citealt{zdz04}). Therefore, the most prominent
feature is that these states are generally observed
during transitions between Hard and Soft states. Within our framework,
they correspond to geometrical situations where $r_J$ is small but remains larger than $r_i$
(Fig.~\ref{diffstate}d). The flux of the outer standard disc is thus
important while the JED is occupying a smaller volume. The consequences on the spectral
shape are not straightforward since the importance of the different
spectral components relative to each other depends on the precise values
of $r_J$ and $\dot m$. Such study is out of the scope of the present
paper and will be detailed elsewhere. \\

The crucial point however is that, in our framework, luminous
intermediate states (the so-called Very High State or VHS) with high $\dot m$ provide the best
conditions for the formation of the ultra-relativistic pair beam, as
described in details in Sect.~\ref{pairbeam}: (1) a high luminosity, (2)
a high energy steep power law spectrum extended up to the $\gamma$-ray
bands and (3) the presence of the MHD jet . The two first characteristics
enable a $\gamma-\gamma$ opacity larger than unity (cf. Eq.~\ref{eqtaugg} of
Sect.~\ref{pairbeam}), while the MHD jet allows to confine the pair beam
and maintain the pair warm, a necessary condition to trigger a pair
runaway process. The total emission would be then dominated by the
explosive behavior of the pairs, with the sudden release of blobs. Each
blob produced in the beam first radiates in X and $\gamma$-ray,
explaining the hard tail present in this state, and then, after a rapid
expansion, produces the optically thin radio emission. This pair beam would also explain the
superluminal ejections observed during this state in different
objects (e.g. \citealt{sob00,han01}). We conjecture that the exact moment where this occurs corresponds to the crossing of the "jet line" recently proposed by \cite{fen04c} (see also
\citealt{cor04}). This corresponds to a transition from the "hard"
intermediate state to the "soft" one.\\ 

The rapid increase of the pair beam pressure in the inner region of the MHD jet, during a
strong outburst, may dramatically perturb the MHD jet production. Indeed,
a huge pair pressure at the axis may enforce the magnetic surfaces to
open dramatically, thereby creating a magnetic compression on the JED
so that no more ejection is feasible. Alternatively,
it is also possible that the racing of the pair process completely wears
out the MHD Poynting flux released by the JED, suppressing the jet
emission or even the jet itself. Whatever occurs (i.e. jet destruction or
jet fading), we expect a suppression of the steady jet emission when a
large outburst sets in. Interestingly, the detailed spectral and timing
study of the radio/X-ray emission of four different black hole binaries
during a major radio outburst \citep{fen04c} shows a weakening and
softening of the X-ray emission as well as a the quenching of the
radio emission after the burst. This is in good agreement with our
expectations since the cooling of the pair beam should indeed results in
a flux decrease and a softening of its spectrum.

%%%%%%%%%%%%%%%%%%%%%%%%%%%%%%%%%%%%%%%%%%%%%%%%%%%
\section{Temporal properties}
%%%%%%%%%%%%%%%%%%%%%%%%%%%%%%%%%%%%%%%%%%%%%%%%%%%

%time scales
Since both SAD and JED are quasi keplerian, the first obvious time scale is the keplerian orbit time, namely 
\begin{equation}
\tau_D (r_J) = \frac{2\pi}{\Omega_k} = 0.1  \left (\frac{m}{10} \right ) \left (\frac{r_J}{50 r_g} \right )^{3/2} \mbox{sec} 
\end{equation}
measured here at the transition radius $r_J$. Since gravity is the dominant force, this time is also the dynamical time involved whenever physical conditions are locally modified within the disc. This time scale is much smaller than the duration of a spectral state or even the transition between two states (but comparable to some timing features, see below).
In fact, the time evolution of BH XrBs requires large variations of the disc accretion rate $\dot m$. The time scales involved, namely rising and decay times but also periodicity, if any, depend therefore on conditions at the outer accretion disc. The inner disc region will thus respond to these variations with its own time scales, which introduces a delay but more importantly a temporal convolution (it acts as a filter). We do dot intend to address the issue of the timing behavior of BH XrBs. As discussed earlier, this requires, within our framework, to take into account the evolution of the large scale magnetic field. Here, we just remark that the presence of these four dynamical components (SAD, JED, MHD jet and pair beam) introduces interesting temporal properties that may be relevant to observations.

Let us assume an increase in $\dot m$ triggered at some outer radius $r_{out} \gg r_J$. This information propagates towards the center with the accretion flow, as a front of increased total pressure. The time scale involved is therefore $\tau_{acc,SAD}(r_{out}) \sim r_{out}/u_r$. If we assume that $r_J$ decreases  (because of a decrease in $\mu$), then the JED (and its associated MHD jets) will progressively disappear. This evolution from a Hard state to an Intermediate or Soft state will be controlled by the advance of this front, namely 
\begin{equation}
\tau_{acc,SAD} \simeq 170\,  \alpha_v^{-1} \left (\frac{\varepsilon}{0.01} \right )^{-2} 
 \left (\frac{m}{10} \right )   \left (\frac{r_J}{50 r_g} \right )^{3/2} \mbox{sec} 
\end{equation}
\noindent If, due to a change in $\dot m$, the radial distribution of the vertical magnetic field needs to be readjusted, then this is done quite fast. Indeed, within the SAD, the time scale for field diffusion is the accretion time scale, since $\tau_{diff} = r^2/\nu_m \sim \tau_{acc}  {\cal R}_m$ with $ {\cal R}_m \sim 1$. Inside the JED, the diffusion time scale is much longer than the accretion time scale because $ {\cal R}_m \sim \varepsilon^{-1}$ \citep{fer95}. However, the accretion time scale inside the JED is itself much shorter than in the SAD (see Eq.~\ref{eq:ms}), so that
\begin{equation}
\tau_{diff,JED} \simeq \alpha_v \, \tau_{acc,SAD}
\end{equation}
Another interesting timing feature is introduced by the MHD jets launched from the JED. Indeed, any adjustment in the disc leads inevitably to a modification of the jet parameters, e.g. the ejection efficiency $\xi$. The time scale for this readjustment can be considered to be of the order of the travel time of the fast MHD waves, namely
\begin{equation}
\tau_{jet}(r) \simeq \int_h^{s_{FM}} \frac{ds}{V_{FM}} 
\end{equation}
where $V_{FM}$ is the speed of the fast magnetosonic wave and the integration is made along a magnetic surface anchored at a disc radius $r$ (from the disc surface $h$ to the fast point $s_{FM}$). 
We computed this time using the MAES solutions of \citet{fer97} and illustrated in his Fig.~6. Solutions 
crossing the fast point right after the Alfv\'en point (larger $\xi$) display $\tau_{jet} (r) \sim \tau_D (r)$ but undergo a recollimation shock right after that. Jet solutions with smaller $\xi$ propagate much farther away and have $\tau_{jet}(r) \sim 10^2 \tau_D (r)$. It is not clear yet if these time scales provide an explanation to some observed timing properties. But the fact that jets are indeed observed is a strong indication that one should take into account their dynamics.

%% QPO 
In our description of the canonical spectral states of XrBs we did not
mention the important issue of quasi-periodic oscillations or
QPOs. Low-frequency (0.1-30 Hz) QPOs in the X-ray bands (2-30 keV) are
indeed observed in the Hard and luminous Intermediate (VHS) states. In the latter state, higher
frequencies (up to 300 Hz) are also present (e.g. \citealt{rem02}, McCR03).
We do not offer yet any precise explanation to these phenomena. However, we note that QPOs are stronger in the hard X-ray bands ($\sim$ 20-30 keV, e.g. \citealt{rod04}) and are correlated with the radio flux \citep{migl05}.  Their emission must then be related to the dynamics involved in the ejection events (both steady and eruptive).

Interestingly, our framework provides a promising environment for the onset of instabilities leading to QPOs. The inner pair beam is an intermittent flow from the inner regions (several $r_i$) and could therefore be responsible for some of the high frequency QPOs. On the other hand, the MHD jet may provide low frequency QPOs. For instance, if some disc material, continuously ejected just outside $r_J$, is failing to become super-Alfv\'enic, then one would expect waves going back and forth between the disc surface and the Alfv\'en surface (located at $r_A$ and $z_A$ in cylindrical coordinates) where a
shock is occurring. A crude estimate of the frequency gives $\nu \sim
V_A/z_A \sim \Omega_* r_A/z_A \sim \Omega_k (r_J)$ since $\Omega_*
r_A\sim V_A$ and $z_A\sim r_A$ in magnetically driven jets \citep{fer97}
and $\Omega_*=\Omega_k(r_J)$ is the rotation rate of the magnetic surface. For  a radius $r_J \sim 50\ r_g$, this gives a $\sim10$ Hz QPO. This clearly deserves further investigation.

%%%%%%%%%%%%%%%%%%%%%%%%%%%%%%%%%%%%%%%%%%%%%%%%%%%
\section{Summary and concluding remarks}
%%%%%%%%%%%%%%%%%%%%%%%%%%%%%%%%%%%%%%%%%%%%%%%%%%%

We present in this paper a new paradigm for the accretion-ejection
properties of Galactic Black Hole X-ray binaries. We assume the existence of a large scale magnetic field of bipolar topology in the innermost disc
regions. Such a field allows for several dynamical phenomena to occur
whose relative importance determine the observed spectral state of the
binary. The dynamical constituents are: (1) an outer standard accretion
disc (SAD) for $r> r_J$, (2) an inner Jet Emitting Disc (JED)
for $r<r_J$ driving (3) a self-collimated non-relativistic
electron-proton surrounding, when adequate conditions are
met, (4) a ultra-relativistic electron-positron beam .
The dynamical properties of each constituent have been thoroughly
analyzed in previous works
(e.g. \citealt{sha73,hen91,fer95,mar97,ren98,sau03,sau04}), but it is the first
time where they are invoked altogether as necessary ingredients to
reproduce the different spectral states of a same object.  

We showed that the various canonical states can be qualitatively explained by varying
{\em independently} the transition radius $r_J$ and the disc accretion rate $\dot m$. In our view, the Quiescent  and Hard states are dominated by non relativistic jet production from the JED, providing henceforth a persistent synchrotron jet emission. The Soft state is obtained when the transition radius $r_J$ becomes smaller than the last marginally stable orbit $r_i$, a SAD is established throughout the whole accretion disc. Intermediate states, between Hard and Soft, are expected to display quite intricate and variable spectral energy distributions. Luminous Intermediate states, obtained during the Hard-to-Soft transitions, are those providing the unique conditions for intermittent pair creation. These pairs give rise to a ultra relativistic beam propagating on the MHD jet axis, explaining both the observed superluminal motions and hard energy tail. \\

% SEDs + comparaisons obs
The dynamical structure presented here (JED, SAD, MHD jet and, occasionally, a
pair beam) seems to be consistent with all available information about
the canonical spectral states of BH XrBs. However, a more quantitative
analysis is critical. In particular, we need to show that the base of the
MHD jet can indeed provide a hot corona with the correct spectral
signature. Then, a precise estimate of the radio/X-ray correlation predicted by
our model and its comparison to observations will be a test of prime
importance for its validity. This is postponed to a future work (Petrucci
et al., in preparation).\\

% hysteresis
In our view, the magnetic flux available at the inner disc regions is a
fundamental and unavoidable ingredient that most probably
varies from one system to another. Since changing the amount of magnetic
flux changes the transition radius $r_J$, the characteristic value of
$\mdot$ (hence luminosity) associated with each spectral state is also
modified. Also, if accreting material is carrying magnetic flux of
opposite direction \citep{tag04}, then this should lead to a major
readjustment of the whole magnetic structure. Clearly, taking into
account the advection of a large scale magnetic field within the disc
introduces a whole new set of variable phenomena.\\

%% Model prediction: gamma-ray emission
Finally, we note that the typical values of the magnetic field required to steadily
launch jets from JEDs, given in Eq.~\ref{Bz}, are consistent with observational estimates \citep{gli99,gne97,gne03}. We expect that BH XrBs should radiate above the MeV range during luminous intermediate states, when pairs are produced. 
A similar proposal has been developed recently by \cite{bos04,bos04b} but it is here a natural
outcome of our model. Very interestingly, such a high energy emission has been recently detected by the HESS instrument in the TeV range \citep{ahar05}. Note also that two microquasars were already detected by EGRET, LS 5039 \citep{par00} and LS I +61 303 \citep[for a recent review]{mas04}, and there is possibly other unidentified galactic EGRET sources \citep{par04}. 
Besides, the $\gamma$-ray emission
of the pair beam can occur further away along the jet at the $\gamma$-ray
photosphere, as proposed for AGNs (e.g. \citealt{mar95}). Noticeably,
$\gamma$-ray spectra of the possible EGRET counterparts seem to exhibit a
break in the MeV range, very similar to that observed in many AGNs. This
break could be explained by the transition from the optically thin X-ray
component to the optically thick, photosphere dominated, $\gamma$-ray
component.
This prediction of $\gamma$-ray emission of microquasars during very high
flaring states could be tested by future GLAST observations.

\begin{acknowledgements}
We thank S. Corbel for a careful reading of the manuscript and T. Belloni for having sent us a draft of its paper before it was completely accepted.
\end{acknowledgements}


\begin{thebibliography}{104}
\expandafter\ifx\csname natexlab\endcsname\relax\def\natexlab#1{#1}\fi

\bibitem[{{Aharonian} {et~al.}(2005){Aharonian}, {Akhperjanian}, {Aye},
  {Bazer-Bachi}, {Beilicke}, {Benbow}, {Berge}, {Berghaus}, {Bernl{\" o}hr},
  {Boisson}, {Bolz}, {Borrel}, {Braun}, {Breitling}, {Brown}, {Gordo},
  {Chadwick}, {Chounet}, {Cornils}, {Costamante}, {Degrange}, {Dickinson},
  {Djannati-Ata{\" i}}, {Drury}, {Dubus}, {Emmanoulopoulos}, {Espigat},
  {Feinstein}, {Fleury}, {Fontaine}, {Fuchs}, {Funk}, {Gallant}, {Giebels},
  {Gillessen}, {Glicenstein}, {Goret}, {Hadjichristidis}, {Hauser},
  {Heinzelmann}, {Henri}, {Hermann}, {Hinton}, {Hofmann}, {Holleran}, {Horns},
  {Jacholkowska}, {de Jager}, {Kh{\' e}lifi}, {Komin}, {Konopelko}, {Latham},
  {Le Gallou}, {Lemi{\` e}re}, {Lemoine-Goumard}, {Leroy}, {Lohse},
  {Marcowith}, {Martin}, {Martineau-Huynh}, {Masterson}, {McComb}, {de
  Naurois}, {Nolan}, {Noutsos}, {Orford}, {Osborne}, {Ouchrif}, {Panter},
  {Pelletier}, {Pita}, {P{\" u}hlhofer}, {Punch}, {Raubenheimer}, {Raue},
  {Raux}, {Rayner}, {Reimer}, {Reimer}, {Ripken}, {Rob}, {Rolland}, {Rowell},
  {Sahakian}, {Saug{\' e}}, {Schlenker}, {Schlickeiser}, {Schuster},
  {Schwanke}, {Siewert}, {Sol}, {Spangler}, {Steenkamp}, {Stegmann},
  {Tavernet}, {Terrier}, {Th{\' e}oret}, {Tluczykont}, {Vasileiadis}, {Venter},
  {Vincent}, {V{\" o}lk}, \& {Wagner}}]{ahar05}
{Aharonian}, F., {Akhperjanian}, A.~G., {Aye}, K.-M., {et~al.} 2005, Science,
  309, 746

\bibitem[{{Balbus} \& {Hawley}(1991)}]{bal91}
{Balbus}, S.~A. \& {Hawley}, J.~F. 1991, \apj, 376, 214

\bibitem[{{Belloni} {et~al.}(2005){Belloni}, {Homan}, P., M., E., W.H.G., J.M.,
  \& M.}]{bel05}
{Belloni}, T., {Homan}, J., P., C., {et~al.} 2005, astro-ph/0504577

\bibitem[{{Belloni} {et~al.}(1997){Belloni}, {Mendez}, {King}, {van der Klis},
  \& {van Paradijs}}]{bel97}
{Belloni}, T., {Mendez}, M., {King}, A.~R., {van der Klis}, M., \& {van
  Paradijs}, J. 1997, \apjl, 479, L145+

\bibitem[{{Beloborodov}(1999)}]{bel99}
{Beloborodov}, A.~M. 1999, in ASP Conf. Ser. 161: High Energy Processes in
  Accreting Black Holes, 295--+

\bibitem[{{Blandford} \& {Begelman}(1999)}]{bla99}
{Blandford}, R.~D. \& {Begelman}, M.~C. 1999, \mnras, 303, L1

\bibitem[{{Blandford} \& {Begelman}(2004)}]{bla04}
{Blandford}, R.~D. \& {Begelman}, M.~C. 2004, \mnras, 349, 68

\bibitem[{{Bogovalov} \& {Tsinganos}(2001)}]{bog01}
{Bogovalov}, S. \& {Tsinganos}, K. 2001, \mnras, 325, 249

\bibitem[{{Bogovalov}(2001)}]{bog01b}
{Bogovalov}, S.~V. 2001, \aap, 371, 1155

\bibitem[{{Bosch-Ramon} \& {Paredes}(2004{\natexlab{a}})}]{bos04b}
{Bosch-Ramon}, V. \& {Paredes}, J.~M. 2004{\natexlab{a}}, \aap, 417, 1075

\bibitem[{{Bosch-Ramon} \& {Paredes}(2004{\natexlab{b}})}]{bos04}
{Bosch-Ramon}, V. \& {Paredes}, J.~M. 2004{\natexlab{b}}, \aap, 425, 1069

\bibitem[{{Buxton} \& {Bailyn}(2004)}]{bux04}
{Buxton}, M.~M. \& {Bailyn}, C.~D. 2004, \apj, 615, 880

\bibitem[{{Casse} \& {Ferreira}(2000{\natexlab{a}})}]{cas00a}
{Casse}, F. \& {Ferreira}, J. 2000{\natexlab{a}}, \aap, 353, 1115

\bibitem[{{Casse} \& {Ferreira}(2000{\natexlab{b}})}]{cas00b}
{Casse}, F. \& {Ferreira}, J. 2000{\natexlab{b}}, \aap, 361, 1178

\bibitem[{{Choudhury} {et~al.}(2003){Choudhury}, {Rao}, {Vadawale}, \&
  {Jain}}]{cho03}
{Choudhury}, M., {Rao}, A.~R., {Vadawale}, S.~V., \& {Jain}, A.~K. 2003, \apj,
  593, 452

\bibitem[{{Corbel} {et~al.}(2004){Corbel}, {Fender}, {Tomsick}, {Tzioumis}, \&
  {Tingay}}]{cor04}
{Corbel}, S., {Fender}, R.~P., {Tomsick}, J.~A., {Tzioumis}, A.~K., \&
  {Tingay}, S. 2004, \apj, 617, 1272

\bibitem[{{Corbel} {et~al.}(2000){Corbel}, {Fender}, {Tzioumis}, {Nowak},
  {McIntyre}, {Durouchoux}, \& {Sood}}]{cor00}
{Corbel}, S., {Fender}, R.~P., {Tzioumis}, A.~K., {et~al.} 2000, \aap, 359, 251

\bibitem[{{Corbel} {et~al.}(2003){Corbel}, {Nowak}, {Fender}, {Tzioumis}, \&
  {Markoff}}]{cor03}
{Corbel}, S., {Nowak}, M.~A., {Fender}, R.~P., {Tzioumis}, A.~K., \& {Markoff},
  S. 2003, \aap, 400, 1007

\bibitem[{{Dhawan} {et~al.}(2000){Dhawan}, {Mirabel}, \&
  {Rodr{\'{\i}}guez}}]{dha00}
{Dhawan}, V., {Mirabel}, I.~F., \& {Rodr{\'{\i}}guez}, L.~F. 2000, \apj, 543,
  373

\bibitem[{{Esin} {et~al.}(2001){Esin}, {McClintock}, {Drake}, {Garcia},
  {Haswell}, {Hynes}, \& {Muno}}]{esi01}
{Esin}, A.~A., {McClintock}, J.~E., {Drake}, J.~J., {et~al.} 2001, \apj, 555,
  483

\bibitem[{{Esin} {et~al.}(1997){Esin}, {McClintock}, \& {Narayan}}]{esi97}
{Esin}, A.~A., {McClintock}, J.~E., \& {Narayan}, R. 1997, \apj, 489, 865

\bibitem[{{Esin} {et~al.}(1998){Esin}, {Narayan}, {Cui}, {Grove}, \&
  {Zhang}}]{esi98}
{Esin}, A.~A., {Narayan}, R., {Cui}, W., {Grove}, J.~E., \& {Zhang}, S.-N.
  1998, \apj, 505, 854

\bibitem[{{Falcke} \& {Biermann}(1995)}]{fal95a}
{Falcke}, H. \& {Biermann}, P.~L. 1995, \aap, 293, 665

\bibitem[{{Falcke} {et~al.}(2004){Falcke}, {K{\" o}rding}, \&
  {Markoff}}]{fal04}
{Falcke}, H., {K{\" o}rding}, E., \& {Markoff}, S. 2004, \aap, 414, 895

\bibitem[{{Fender} \& {Belloni}(2004)}]{fen04b}
{Fender}, R. \& {Belloni}, T. 2004, \araa, 42, 317

\bibitem[{{Fender} {et~al.}(1999){Fender}, {Corbel}, {Tzioumis}, {McIntyre},
  {Campbell-Wilson}, {Nowak}, {Sood}, {Hunstead}, {Harmon}, {Durouchoux}, \&
  {Heindl}}]{fen99}
{Fender}, R., {Corbel}, S., {Tzioumis}, T., {et~al.} 1999, \apjl, 519, L165

\bibitem[{{Fender} {et~al.}(2004){Fender}, {Belloni}, \& {Gallo}}]{fen04c}
{Fender}, R.~P., {Belloni}, T.~M., \& {Gallo}, E. 2004, \mnras, 355, 1105

\bibitem[{{Fender} {et~al.}(2003){Fender}, {Gallo}, \& {Jonker}}]{fen03}
{Fender}, R.~P., {Gallo}, E., \& {Jonker}, P.~G. 2003, \mnras, 343, L99

\bibitem[{{Ferreira}(1997)}]{fer97}
{Ferreira}, J. 1997, \aap, 319, 340

\bibitem[{{Ferreira}(2002)}]{fer02}
{Ferreira}, J. 2002, in "Star Formation and the Physics of Young Stars",
  J.~Bouvier and J.-P.~Zahn (Eds), EAS Publications Series, astro-ph/0311621,
  3, 229

\bibitem[{{Ferreira} \& {Casse}(2004)}]{fer04}
{Ferreira}, J. \& {Casse}, F. 2004, \apjl, 601, L139

\bibitem[{{Ferreira} \& {Pelletier}(1993)}]{fer93a}
{Ferreira}, J. \& {Pelletier}, G. 1993, \aap, 276, 625+

\bibitem[{{Ferreira} \& {Pelletier}(1995)}]{fer95}
{Ferreira}, J. \& {Pelletier}, G. 1995, \aap, 295, 807+

\bibitem[{{Frontera} {et~al.}(2001){Frontera}, {Palazzi}, {Zdziarski},
  {Haardt}, {Perola}, {Chiappetti}, {Cusumano}, {Dal Fiume}, {Del Sordo},
  {Orlandini}, {Parmar}, {Piro}, {Santangelo}, {Segreto}, {Treves}, \&
  {Trifoglio}}]{fro01}
{Frontera}, F., {Palazzi}, E., {Zdziarski}, A.~A., {et~al.} 2001, \apj, 546,
  1027

\bibitem[{{Galeev} {et~al.}(1979){Galeev}, {Rosner}, \& {Vaiana}}]{gal79}
{Galeev}, A.~A., {Rosner}, R., \& {Vaiana}, G.~S. 1979, \apj, 229, 318

\bibitem[{{Gallo} {et~al.}(2004){Gallo}, {Corbel}, {Fender}, {Maccarone}, \&
  {Tzioumis}}]{gal04}
{Gallo}, E., {Corbel}, S., {Fender}, R.~P., {Maccarone}, T.~J., \& {Tzioumis},
  A.~K. 2004, \mnras, 347, L52

\bibitem[{{Gallo} {et~al.}(2005){Gallo}, {Fender}, \& {Hynes}}]{gal05}
{Gallo}, E., {Fender}, R.~P., \& {Hynes}, R.~I. 2005, \mnras, 356, 1017

\bibitem[{{Gallo} {et~al.}(2003){Gallo}, {Fender}, \& {Pooley}}]{gal03}
{Gallo}, E., {Fender}, R.~P., \& {Pooley}, G.~G. 2003, \mnras, 344, 60

\bibitem[{{Gilfanov} {et~al.}(1999){Gilfanov}, {Churazov}, \&
  {Revnivtsev}}]{gil99}
{Gilfanov}, M., {Churazov}, E., \& {Revnivtsev}, M. 1999, \aap, 352, 182

\bibitem[{{Gliozzi} {et~al.}(1999){Gliozzi}, {Bodo}, \& {Ghisellini}}]{gli99}
{Gliozzi}, M., {Bodo}, G., \& {Ghisellini}, G. 1999, \mnras, 303, L37

\bibitem[{{Gnedin} {et~al.}(2003){Gnedin}, {Borisov}, {Natsvlishvili},
  {Piotrovich}, \& {Silant'ev}}]{gne03}
{Gnedin}, Y.~N., {Borisov}, N., {Natsvlishvili}, T., {Piotrovich}, M., \&
  {Silant'ev}, N. 2003, in

\bibitem[{{Gnedin} \& {Natsvlishvili}(1997)}]{gne97}
{Gnedin}, Y.~N. \& {Natsvlishvili}, T.~M. 1997, in Stellar Magnetic Fields,
  Proceedings of the International Conference, held in the Special
  Astrophysical Observatory of the Russian AS, May 13-18, 1996, Eds.: Yu.
  Glagolevskij, I. Romanyuk, Special Astrophysical Observatory Press, p.
  40-54., 40--54

\bibitem[{{Grove} {et~al.}(1998){Grove}, {Johnson}, {Kroeger}, {McNaron-Brown},
  {Skibo}, \& {Phlips}}]{gro98}
{Grove}, J.~E., {Johnson}, W.~N., {Kroeger}, R.~A., {et~al.} 1998, \apj, 500,
  899

\bibitem[{{Haardt}(1993)}]{haa93}
{Haardt}, F. 1993, \apj, 413, 680

\bibitem[{{Hameury} {et~al.}(1997){Hameury}, {Lasota}, {McClintock}, \&
  {Narayan}}]{ham97}
{Hameury}, J.-M., {Lasota}, J.-P., {McClintock}, J.~E., \& {Narayan}, R. 1997,
  \apj, 489, 234

\bibitem[{{Hannikainen} {et~al.}(2001){Hannikainen}, {Campbell-Wilson},
  {Hunstead}, {McIntyre}, {Lovell}, {Reynolds}, {Tzioumis}, \& {Wu}}]{han01}
{Hannikainen}, D., {Campbell-Wilson}, D., {Hunstead}, R., {et~al.} 2001,
  Astrophysics and Space Science Supplement, 276, 45

\bibitem[{{Heinz}(2004)}]{hei04}
{Heinz}, S. 2004, \mnras, 355, 835

\bibitem[{{Heinz} \& {Sunyaev}(2003)}]{hei03}
{Heinz}, S. \& {Sunyaev}, R.~A. 2003, \mnras, 343, L59

\bibitem[{{Henri} \& {Pelletier}(1991)}]{hen91}
{Henri}, G. \& {Pelletier}, G. 1991, \apjl, 383, L7

\bibitem[{{Heyvaerts} \& {Priest}(1989)}]{hey89a}
{Heyvaerts}, J.~F. \& {Priest}, E.~R. 1989, \aap, 216, 230

\bibitem[{{K{\" o}nigl}(2004)}]{kon04}
{K{\" o}nigl}, A. 2004, \apj, 617, 1267

\bibitem[{{Kalemci} {et~al.}(2005){Kalemci}, {Tomsick}, {Buxton}, {Rothschild},
  {Pottschmidt}, {Corbel}, {Brocksopp}, \& {Kaaret}}]{kale05}
{Kalemci}, E., {Tomsick}, J.~A., {Buxton}, M.~M., {et~al.} 2005, \apj, 622, 508

\bibitem[{{King} {et~al.}(2004){King}, {Pringle}, {West}, \& {Livio}}]{kin04}
{King}, A.~R., {Pringle}, J.~E., {West}, R.~G., \& {Livio}, M. 2004, \mnras,
  348, 111

\bibitem[{{Lasota} {et~al.}(1996){Lasota}, {Narayan}, \& {Yi}}]{las96}
{Lasota}, J.-P., {Narayan}, R., \& {Yi}, I. 1996, \aap, 314, 813

\bibitem[{{Livio} {et~al.}(2003){Livio}, {Pringle}, \& {King}}]{liv03}
{Livio}, M., {Pringle}, J.~E., \& {King}, A.~R. 2003, \apj, 593, 184

\bibitem[{{Maccarone}(2003)}]{mac03b}
{Maccarone}, T.~J. 2003, \aap, 409, 697

\bibitem[{{Mahadevan}(1997)}]{mah97}
{Mahadevan}, R. 1997, \apj, 477, 585

\bibitem[{{Malzac} {et~al.}(2001){Malzac}, {Beloborodov}, \&
  {Poutanen}}]{mal01}
{Malzac}, J., {Beloborodov}, A.~M., \& {Poutanen}, J. 2001, \mnras, 326, 417

\bibitem[{{Marcowith} {et~al.}(1995){Marcowith}, {Henri}, \&
  {Pelletier}}]{mar95}
{Marcowith}, A., {Henri}, G., \& {Pelletier}, G. 1995, \mnras, 277, 681

\bibitem[{{Marcowith} {et~al.}(1998){Marcowith}, {Henri}, \& {Renaud}}]{mar98}
{Marcowith}, A., {Henri}, G., \& {Renaud}, N. 1998, \aap, 331, L57

\bibitem[{{Marcowith} {et~al.}(1997){Marcowith}, {Pelletier}, \&
  {Henri}}]{mar97}
{Marcowith}, A., {Pelletier}, G., \& {Henri}, G. 1997, \aap, 323, 271

\bibitem[{{Markoff}(2004)}]{mark04b}
{Markoff}, S. 2004, in X-ray Binaries to Quasars: Black Hole Accretion on All
  Mass Scales, ed. T. J. Maccarone, R. P. Fender, and L. C. Ho

\bibitem[{{Markoff} {et~al.}(2001){Markoff}, {Falcke}, \& {Fender}}]{mark01}
{Markoff}, S., {Falcke}, H., \& {Fender}, R. 2001, \aap, 372, L25

\bibitem[{{Markoff} {et~al.}(2003){Markoff}, {Nowak}, {Corbel}, {Fender}, \&
  {Falcke}}]{mark03}
{Markoff}, S., {Nowak}, M., {Corbel}, S., {Fender}, R., \& {Falcke}, H. 2003,
  \aap, 397, 645

\bibitem[{{Markoff} \& {Nowak}(2004)}]{mark04}
{Markoff}, S. \& {Nowak}, M.~A. 2004, \apj, 609, 972

\bibitem[{{Massi} {et~al.}(2004){Massi}, M., J.M., {Garrington}, {Peracaula},
  \& {Marti}}]{mas04}
{Massi}, J.~M., M., R., J.M., P., {et~al.} 2004, in AIP Proceedings Series
  astro-ph/0410504

\bibitem[{{McClintock} \& {Remillard}(2003)}]{mcc03}
{McClintock}, J.~E. \& {Remillard}, R.~A. 2003, astro-ph/0306213 (McCR03)

\bibitem[{{Merloni} \& {Fabian}(2002)}]{mer02}
{Merloni}, A. \& {Fabian}, A.~C. 2002, \mnras, 332, 165

\bibitem[{{Merloni} {et~al.}(2003){Merloni}, {Heinz}, \& {di Matteo}}]{mer03}
{Merloni}, A., {Heinz}, S., \& {di Matteo}, T. 2003, \mnras, 345, 1057

\bibitem[{{Migliari} {et~al.}(2005){Migliari}, {Fender}, \& M.}]{migl05}
{Migliari}, S., {Fender}, R., \& M., v. 2005, astro-ph/0507223

\bibitem[{{Mirabel} \& {Rodriguez}(1998)}]{mir98}
{Mirabel}, I.~F. \& {Rodriguez}, L.~F. 1998, \nat, 392, 673

\bibitem[{{Mirabel} \& {Rodr{\'{\i}}guez}(1999)}]{mir99}
{Mirabel}, I.~F. \& {Rodr{\'{\i}}guez}, L.~F. 1999, \araa, 37, 409

\bibitem[{{Narayan} {et~al.}(1996){Narayan}, {McClintock}, \& {Yi}}]{nar96}
{Narayan}, R., {McClintock}, J.~E., \& {Yi}, I. 1996, \apj, 457, 821

\bibitem[{{Nowak} {et~al.}(2002){Nowak}, {Wilms}, \& {Dove}}]{now02}
{Nowak}, M.~A., {Wilms}, J., \& {Dove}, J.~B. 2002, \mnras, 332, 856

\bibitem[{{O'Dell}(1981)}]{ode81}
{O'Dell}, S.~L. 1981, \apjl, 243, L147

\bibitem[{{Paredes}(2004)}]{par04}
{Paredes}, J.~M. 2004, in V Microquasar Workshop, Beijing, June 2004
  astro-ph/0409226

\bibitem[{{Paredes} {et~al.}(2000){Paredes}, {Mart{\'{\i}}}, {Rib{\' o}}, \&
  {Massi}}]{par00}
{Paredes}, J.~M., {Mart{\'{\i}}}, J., {Rib{\' o}}, M., \& {Massi}, M. 2000,
  Science, 288, 2340

\bibitem[{{Pelletier}(2004)}]{pel04}
{Pelletier}, G. 2004, in Dynamics and dissipation in electromagnetically
  dominated media" (Nova Science) edited by M. Lyutikov (astro-ph/0405113)

\bibitem[{{Pelletier} \& {Roland}(1989)}]{pel89}
{Pelletier}, G. \& {Roland}, J. 1989, \aap, 224, 24

\bibitem[{{Pelletier} \& {Sol}(1992)}]{pel92}
{Pelletier}, G. \& {Sol}, H. 1992, \mnras, 254, 635

\bibitem[{{Pelletier} {et~al.}(1988){Pelletier}, {Sol}, \& {Asseo}}]{pel88b}
{Pelletier}, G., {Sol}, H., \& {Asseo}, E. 1988, \pra, 38, 2552

\bibitem[{{Petrucci} {et~al.}(2000){Petrucci}, {Haardt}, {Maraschi}, {Grandi},
  {Matt}, {Nicastro}, {Piro}, {Perola}, \& {De Rosa}}]{pet00}
{Petrucci}, P.~O., {Haardt}, F., {Maraschi}, L., {et~al.} 2000, \apj, 540, 131

\bibitem[{{Phinney}(1987)}]{phin87}
{Phinney}, E.~S. 1987, in Superluminal Radio Sources, 301--305

\bibitem[{{Pietrini} \& {Krolik}(1995)}]{kro95}
{Pietrini}, P. \& {Krolik}, J.~H. 1995, \apj, 447, 526

\bibitem[{{Poutanen} \& {Svensson}(1996)}]{pou96}
{Poutanen}, J. \& {Svensson}, R. 1996, \apj, 470, 249

\bibitem[{{Remillard} {et~al.}(2002){Remillard}, {Sobczak}, {Muno}, \&
  {McClintock}}]{rem02}
{Remillard}, R.~A., {Sobczak}, G.~J., {Muno}, M.~P., \& {McClintock}, J.~E.
  2002, \apj, 564, 962

\bibitem[{{Renaud} \& {Henri}(1998)}]{ren98}
{Renaud}, N. \& {Henri}, G. 1998, \mnras, 300, 1047

\bibitem[{{Robertson} \& {Leiter}(2004)}]{rob04}
{Robertson}, S.~L. \& {Leiter}, D.~J. 2004, \mnras, 350, 1391

\bibitem[{{Rodriguez} {et~al.}(2004){Rodriguez}, {Corbel}, {Hannikainen},
  {Belloni}, {Paizis}, \& {Vilhu}}]{rod04}
{Rodriguez}, J., {Corbel}, S., {Hannikainen}, D.~C., {et~al.} 2004, \apj, 615,
  416

\bibitem[{{Romanova} {et~al.}(1998){Romanova}, {Ustyugova}, {Koldoba},
  {Chechetkin}, \& {Lovelace}}]{rom98}
{Romanova}, M.~M., {Ustyugova}, G.~V., {Koldoba}, A.~V., {Chechetkin}, V.~M.,
  \& {Lovelace}, R.~V.~E. 1998, \apj, 500, 703

\bibitem[{{Saug{\' e}} \& {Henri}(2003)}]{sau03}
{Saug{\' e}}, L. \& {Henri}, G. 2003, New Astronomy Review, 47, 529

\bibitem[{{Saug{\' e}} \& {Henri}(2004)}]{sau04}
{Saug{\' e}}, L. \& {Henri}, G. 2004, \apj, 616, 136

\bibitem[{{Shakura} \& {Sunyaev}(1973)}]{sha73}
{Shakura}, N.~I. \& {Sunyaev}, R.~A. 1973, \aap, 24, 337

\bibitem[{{Sidoli} \& {Mereghetti}(2002)}]{sid02}
{Sidoli}, L. \& {Mereghetti}, S. 2002, \aap, 388, 293

\bibitem[{{Sobczak} {et~al.}(2000){Sobczak}, {McClintock}, {Remillard}, {Cui},
  {Levine}, {Morgan}, {Orosz}, \& {Bailyn}}]{sob00}
{Sobczak}, G.~J., {McClintock}, J.~E., {Remillard}, R.~A., {et~al.} 2000, \apj,
  544, 993

\bibitem[{{Sol} {et~al.}(1989){Sol}, {Pelletier}, \& {Asseo}}]{sol89}
{Sol}, H., {Pelletier}, G., \& {Asseo}, E. 1989, \mnras, 237, 411

\bibitem[{{Stirling} {et~al.}(2001){Stirling}, {Spencer}, {de la Force},
  {Garrett}, {Fender}, \& {Ogley}}]{sti01}
{Stirling}, A.~M., {Spencer}, R.~E., {de la Force}, C.~J., {et~al.} 2001,
  \mnras, 327, 1273

\bibitem[{{Stone} {et~al.}(1996){Stone}, {Hawley}, {Gammie}, \&
  {Balbus}}]{sto96}
{Stone}, J.~M., {Hawley}, J.~F., {Gammie}, C.~F., \& {Balbus}, S.~A. 1996,
  \apj, 463, 656

\bibitem[{{Tagger} {et~al.}(2004){Tagger}, {Varni{\` e}re}, {Rodriguez}, \&
  {Pellat}}]{tag04}
{Tagger}, M., {Varni{\` e}re}, P., {Rodriguez}, J., \& {Pellat}, R. 2004, \apj,
  607, 410

\bibitem[{{Tanaka} \& {Lewin}(1995)}]{tan95c}
{Tanaka}, Y. \& {Lewin}, W. H.~G. 1995, in X-Ray Binaries, ed. W. H. G. Lewin,
  J. van Paradijs, and E. P. J. van den Heuvel (Cambridge: Cambridge Univ.
  Press), 126

\bibitem[{{Vadawale} {et~al.}(2001){Vadawale}, {Rao}, \& {Chakrabarti}}]{vad01}
{Vadawale}, S.~V., {Rao}, A.~R., \& {Chakrabarti}, S.~K. 2001, \aap, 372, 793

\bibitem[{{Zdziarski} {et~al.}(2004){Zdziarski}, {Gierli{\' n}ski},
  {Miko{\l}ajewska}, {Wardzi{\' n}ski}, {Smith}, {Alan Harmon}, \&
  {Kitamoto}}]{zdz04b}
{Zdziarski}, A.~A., {Gierli{\' n}ski}, M., {Miko{\l}ajewska}, J., {et~al.}
  2004, \mnras, 351, 791

\bibitem[{{Zdziarski} \& {Gierlinski}(2003)}]{zdz04}
{Zdziarski}, A.~A. \& {Gierlinski}, M. 2003, in Proceedings of "Stellar-mass,
  intermediate-mass, and supermassive black holes", Kyoto astro-ph/0403683
  (ZG04), 99--119

\bibitem[{{Zdziarski} {et~al.}(1999){Zdziarski}, {Lubinski}, \&
  {Smith}}]{zdz99}
{Zdziarski}, A.~A., {Lubinski}, P., \& {Smith}, D.~A. 1999, \mnras, 303, L11

\end{thebibliography}
\end{document}